%% file: gcl_absorption.tex
\documentclass[twocolumn]{aastex62} 
\usepackage{lineno}

\input{preamble.tex}

\shorttitle{Absorption toward the GCL}
\shortauthors{Hurley-Walker et al.}

\begin{document}

\title{Low-frequency absorption and radio recombination line features of the Galactic Center Lobe}

\author[0000-0002-5119-4808]{Natasha Hurley-Walker}
\affiliation{International Centre for Radio Astronomy Research, Curtin University, Bentley, WA 6102, Australia}

\author[0000-0001-8800-1793]{L.~D.~Anderson}
\affiliation{Department of Physics and Astronomy, West Virginia University, Morgantown, WV 26506}
\affiliation{Adjunct Astronomer at the Green Bank Observatory, P.O. Box 2, Green Bank, WV 24944}
\affiliation{Center for Gravitational Waves and Cosmology, West Virginia University, Chestnut Ridge Research Building, Morgantown, WV 26505}

\author[0000-0001-8061-216X]{M.~Luisi}
\affiliation{Department of Physics, Westminster College, New Wilmington, PA 16172}
\affiliation{Center for Gravitational Waves and Cosmology, West Virginia University, Chestnut Ridge Research Building, Morgantown, WV 26505}

\author[0000-0003-2730-957X]{N.~M.~McClure-Griffiths}
\affiliation{Research School of Astronomy \& Astrophysics, Australian National University, Canberra 2600 ACT Australia}

\author[0000-0002-8109-2642]{Robert A. Benjamin}
\affil{University of Wisconsin-Whitewater, 800 W. Main St, Whitewater, WI 53190, USA}

\author[0000-0002-0631-7514]{Michael A. Kuhn}
\affiliation{Department of Astronomy, California Institute of Technology, Pasadena, CA, 91125, USA}

\author[0000-0002-4727-7619]{Dylan~J.~Linville}
\affiliation{Department of Physics and Astronomy, West Virginia University, Morgantown, WV 26506}
\affiliation{Center for Gravitational Waves and Cosmology, West Virginia University, Chestnut Ridge Research Building, Morgantown, WV 26505}

\author[0000-0002-1311-8839]{B. Liu}
\affiliation{CAS Key Laboratory of FAST, National Astronomical Observatories, Chinese Academy of Sciences, Beijing 100101, People's Republic of China}

\author[0000-0002-2250-730X]{Catherine Zucker}
\affiliation{Harvard Astronomy, Harvard-Smithsonian Center for Astrophysics, 60 Garden St, Cambridge, MA, 02138, USA}

\correspondingauthor{N.~Hurley-Walker}
\email{nhw@icrar.org}

\begin{abstract}
The Galactic center lobe (GCL) is a $\sim\!1\degree$ object located north of the Galactic center.  In the mid-infrared (MIR), the GCL appears as two 8.0\,\micron\ filaments that roughly define an ellipse.  There is strong 24\,\micron\ and radio continuum emission in the interior of the ellipse.  Due to its morphology and location in the sky, previous authors have argued that the GCL is created by outflows from star formation in the central molecular zone or by activity of the central black hole Sgr~A$^*$. We present images of the GCL from the GaLactic and Extragalactic All-sky Murchison Widefield Array survey in radio continuum that show thermal absorption against the Galactic center, incompatible with an interpretation of synchrotron self-absorption. Estimates of the cosmic ray emissivity in this direction allow us to place a distance constraint on the GCL. To be consistent
with standard emissivity assumptions, the GCL would be located 2\,kpc
away. At a distance of 8\,kpc, the synchrotron background emissivity is enhanced by $\sim75$\,\%  in the direction of the GCL. We also present radio recombination line data from the Green Bank Telescope that constrains the electron temperature and line widths in this region, which are also more explicable if the GCL lies relatively close.
\end{abstract}

\keywords{\hii\ regions (694), Radio continuum emission (1340), 
Interstellar line emission (844), Interstellar medium (847), Interstellar absorption (831)}

\section{Introduction}

The center of the Milky Way is home to intense star formation and a central black hole, Sgr\,A$^*$, that has a mass of $4\times 10^6\,\msun$ \citep{gillessen09}.  There is little evidence that the central black hole is actively accreting. Emission from the ``{\it Fermi} bubbles'' \citep{su10}, bipolar $\gamma$-ray structures that extend some 50\degree\ above and below the plane, can be interpreted as the signature of past accretion events.
An actively accreting black hole would power outflows that could be traced, for example, at radio wavelengths. 

North of the Galactic center there exists a structure known as the ``Galactic Center Lobe'' (GCL), sometimes called the ``Galactic Center Omega Lobe.''  
The GCL is about $1\degree$ across, and roughly elliptical.  If located in the Galactic center at a distance of 8.2\,\kpc\ \citep{abuter19}, the GCL would have a diameter of $\sim\!140\,\pc$.  At mid-infrared (MIR) wavelengths, the GCL appears as two 8.0\,\micron\ filaments, with the region between them filled with 24\,\micron\ emission; thermal radio emission from plasma fills the space between the two 8.0\,\micron\ filaments.

The GCL is a conspicuous radio continuum structure that is seen in absorption at low radio frequencies.  \citet{sofue84} first noted the existence of the GCL in 10\,\ghz\ continuum data \citep[see also][]{sofue85}.  They suggested that it was thermally emitting and hypothesized that it was created by an energetic outflow from Sgr~A$^*$.  The GCL is also readily apparent in the 1.4 and 2.7\,\ghz\ continuum data of \citet{pohl92}.  The GCL may have a connection to large-scale polarized radio continuum emission \citep{caretti13}.  \citet{tsuboi85} found that the eastern half of the GCL is strongly polarized, but the western half is not, which was supported by \citet{haynes92}.
\citet{brogan03} found free-free absorption of the GCL using multi-configuration Very Large Array (VLA) 74\,\mhz\ data.  The western part of the GCL is more strongly absorbed than the eastern half.

Multiple authors have suggested that the GCL is created by activity in the Galactic center, either due to Sgr~A$^*$ \citep[e.g.,][]{uchida85,kassim86} or due to star formation in the Central Molecular Zone \citep[e.g.][]{law09}.
\citet{law09} detected radio recombination line (RRL) emission from the GCL using the Green Bank Telescope (GBT), which proved the existence of thermal emission. 
They found that the electron temperature, gas pressure and morphology of the GCL are consistent with it being located in the Galactic center.
The GCL RRL line widths are narrower than those of most Galactic \hii\ regions, which implies low turbulence, low electron temperatures, or both.  \citet{law09} estimated that three large star clusters in the Galactic center region could account for the ionization of the GCL.

If the GCL is created from activity in the Galactic center, there may be a complementary feature south of the Galactic center.  \citet{bland-hawthorn03} suggested that the GCL is created by winds emanating from Sgr~A$^*$ and identified a feature in {\it Midcourse Space Experiment} (MSX) MIR data located south of the Galactic center. Recently, \citet{heywood19} observed the Galactic center with the MeerKAT radio telescope array at 1.28\,\mhz\ and $6\arcsec$ angular resolution.  They clearly detected the GCL, and found a complementary structure south of the Galactic center \citep[distinct from the structure found by][]{bland-hawthorn03} that they posited is connected to the GCL.  
\citet{heywood19} interpret these features as arising from an explosive, energetic event near Sgr~A$^*$ that occurred a few million years ago.\footnote{We note that the more recent data release by \citet{2022ApJ...925..165H} shows the base of the GCL, but does not provide spectral index measurements of the GCL, nor imaging at higher latitudes.}  There is also X-ray emission seen toward the northern and southern features that further supports the explosion hypothesis \citep{nakashima13,ponti19}.  \citet{heywood19} note that the southern radio continuum emission encloses the X-ray emission.

Arguments that the GCL is foreground to the Galactic center are also found in the literature.  \citet{nagoshi19} detected RRL emission from the GCL using the Yamaguchi 32-m radio telescope and argued that the velocities of the detected thermal ionized gas are inconsistent with Galactic rotation of an object in the Galactic center \citep[although][did find evidence of fast associated molecular gas]{sofue96}. \cite{tsuboi20} showed further evidence that the GCL is foreground to the Galactic center by noting that a section of the GCL is correlated with optical extinction feature while a higher latitude section ($b > 0.8\degree$) is seen in H$\alpha$ emission.  That the GCL is seen in absorption at low radio frequencies is also evidence that it is foreground to the intense synchrotron emission found in the Galactic center \citep[see][]{kassim90, brogan03}.

The data and analysis to date paint a confusing picture for the distance, size, and origin of the GCL. This paper is one of two papers which present new lines of evidence on the nature and distance of the GCL. In this paper, we examine the absorption of low radio frequencies by the structure, and its signature in radio recombination lines, to place constraints on its distance. In the companion paper we compare the mid-infrared and radio properties of the GCL with those of Galactic \hii regions.

\section{Data}
\begin{figure*}
    \includegraphics[width=0.49\textwidth]{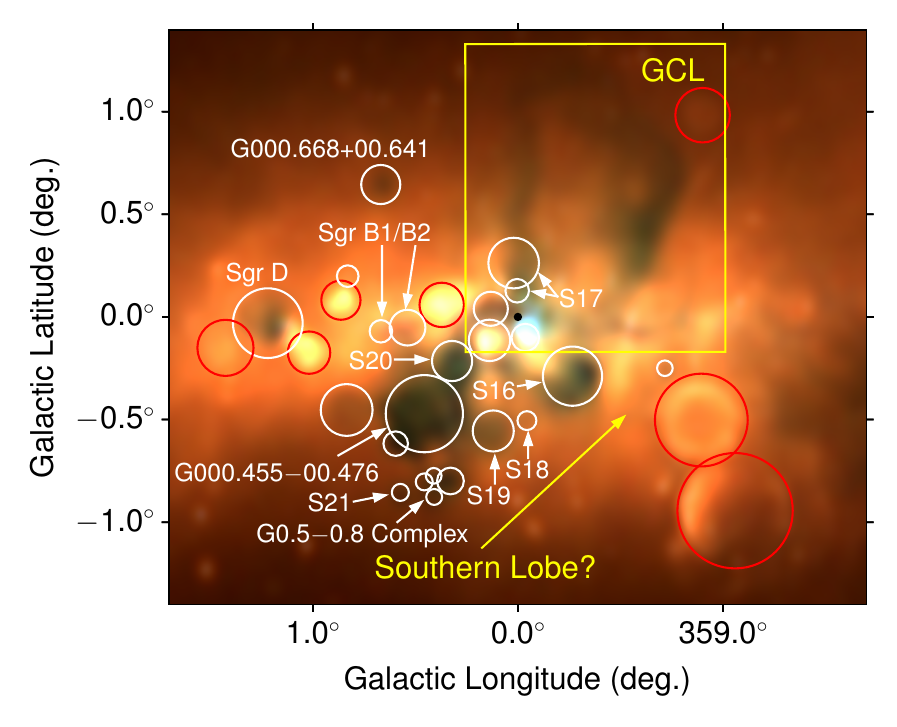}
    \includegraphics[width=0.49\textwidth]{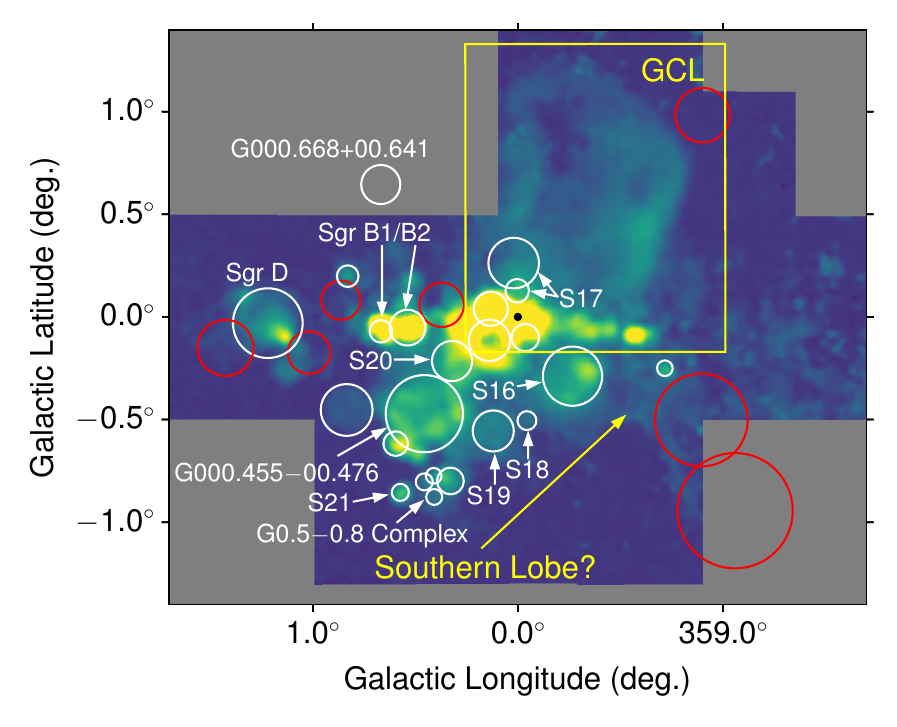}
\caption{Overview of the Galactic center region. Left: Murchison Widefield Array GLEAM three-color image with 72--103\,\mhz\ data in red, 103--134\,\mhz\ data in green, and 139--170\,\mhz\ data in blue. 
Right: Green Bank Telescope GDIGS integrated \hna\ RRL intensity, integrated over $\pm20\,\kms$.  White circles in both panels enclose particularly bright \hii\ regions from the {\it WISE} Catalog of Galactic \hii\ Regions \citep{anderson14} and red circles enclose known supernova remnants \citep{green06}.  The black dot shows the location of Sgr~A$^*$.  The yellow rectangle encloses the GCL and the location of the proposed Southern Lobe is indicated \citep{bland-hawthorn03}.  
    \label{fig:gc}}
\end{figure*}

We use data from two main sources to investigate the emission of the GCL.  We show overview images of the Galactic center from these data in Figure~\ref{fig:gc}.

\subsection{MWA GLEAM data\label{sec:gleam}}

The GCL was imaged by the Murchison Widefield Array \citep[MWA; ][]{2013PASA...30....7T} between 72\,MHz and 231\,MHz in 20 7.68\,\mhz-wide sub-bands as part of the GaLactic and Extragalactic All-sky MWA survey \citep[GLEAM;][]{2015PASA...32...25W,2017MNRAS.464.1146H,2019PASA...36...47H}. GLEAM has a resolution of $\sim\!4-2\arcmin$ over 72--231\,\mhz. The GLEAM data are well-suited to measuring diffuse Galactic emission, as they are sensitive to angular scales of up to $\sim\!30-15\degree$ over the frequency range.
At the low radio frequencies observed by the MWA, the majority of the emission from the Galactic plane is non-thermal synchrotron from cosmic rays interacting with the Galactic magnetic field. 
Figure~\ref{fig:gc} shows the GCL in absorption against the bright Galactic background at frequencies below $\sim\!150\,\mhz$.  The strength of the absorption is correlated with the strength of RRL emission.

\begin{figure*}
    \centering
    \includegraphics[width=\textwidth]{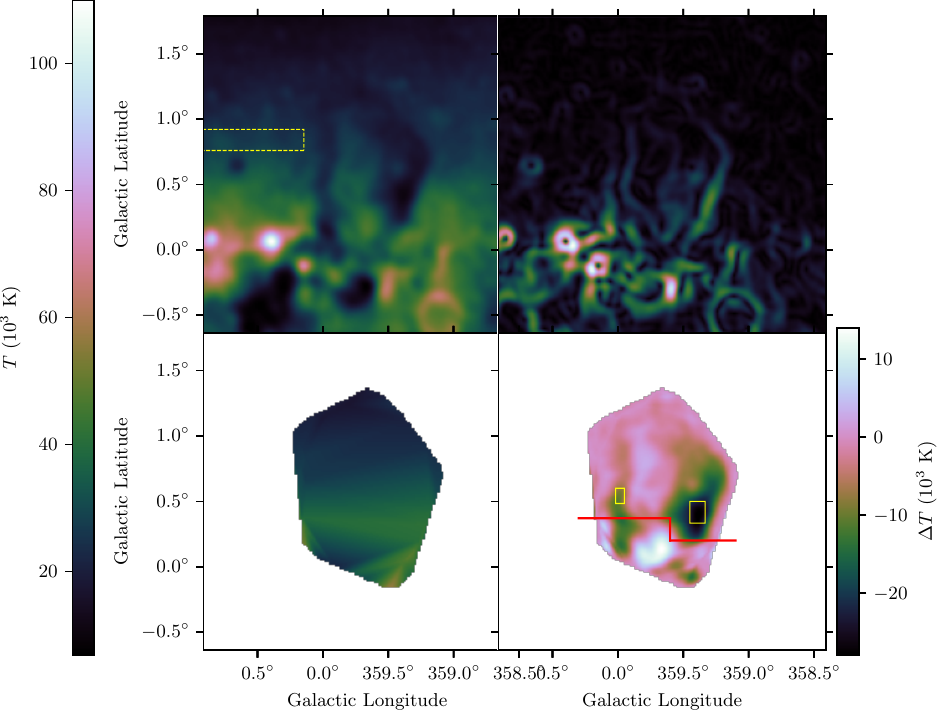}
    \caption{Background estimation and subtraction for the GLEAM 76\,\mhz\ sub-band. The top-left panel is the original image, centered on the GCL. The top-right panel shows the result of edge detection with a Sobel-Feldman filter. The bottom-left panel encloses the area grown by one beamwidth, and shows the background calculated by linear interpolation. The bottom-right panel shows the background-subtracted image. The dashed box in the top-left panel shows a source-free region selected to measure the RMS as an input to the uncertainty calculation in Section~\ref{sec:gleam_abs}. The two yellow boxes in the bottom-right panel show locations chosen to display SEDs in Section~\ref{sec:gleam_abs}. The red lines in the lower right panel indicate the limit of good background subtraction sufficient to perform an absorption analysis (Section~\ref{sec:gleam_abs}).
    \label{fig:sobel}}
\end{figure*}

The GCL lies in an area of complex emission in the low-frequency radio band, so careful background subtraction is required for quantitative analyses. We perform background subtraction individually in each GLEAM sub-band, and show the 76\,\mhz\ analysis in Figure~\ref{fig:sobel}.
First, we regrid and convolve all GLEAM sub-band images to the same pixel grid and resolution of $5\farcm4\times4\farcm4$ 
. The GLEAM maps are natively in units of \jyb and we convert to brightness temperature $T_B$ in \K via the standard equation derived from the Rayleigh-Jeans law:
\begin{equation}
    \frac{T_B}{\K} = \frac{1.222\times10^3} {\theta_\mathrm{maj} \theta_\mathrm{min}}\left(\frac{\nu }{\rm \ghz}\right)^{-2}\left(\frac{S}{\rm \jyb}\right)\,,
\end{equation}
where $\theta_\mathrm{maj}$ and $\theta_\mathrm{min}$ are the major and minor axes of the restoring beam (in radians), respectively. We use these regridded images in \K\ for all GLEAM analyses.

We use a Sobel-Feldman operator \citep{sobel,1973pcsa.book.....D,FreemanMV3D1990} implemented in the \textsc{Python} package \textsc{scipy} to detect the edges of the GCL in the 76\,\mhz\ image.  We then expand the edges by one beamwidth ($5\arcmin$), and make manual alterations to the northern border where the signal is low, creating an outline mask of the GCL. We select an area one beamwidth wide around this path to estimate the background.  

For each GLEAM sub-band, we linearly interpolate the background path pixel values to generate a two-dimensional map of the background level across the GCL. We subtract the background maps from each sub-band image to produce background-subtracted images of the GCL, which are then used for analysis of the absorption (Section~\ref{sec:gleam_abs}).

\subsection{GBT RRL data}

We use data from the Green Bank Telescope (GBT) Diffuse Ionized Gas Survey \citep[GDIGS;][]{2021ApJS..254...28A}, which is a fully-sampled C-band (4-8\,\ghz) RRL survey of the inner Galactic plane using the GBT.  RRLs trace the emission from thermal plasma.  
GDIGS simultaneously measures the emission from 22 Hn$\alpha$ RRL transitions, ranging from $n=95$ to 117, 15 of which are averaged to create final Hn$\alpha$ maps of RRL emission across the Galactic plane at 2\farcm65 resolution.

We fit Gaussian components to data from each GDIGS spaxel that has emission of at least 3$\sigma$ using GaussPy+ \citep{riener19}, where $\sigma$ is the spectral rms computed individually from the line-free portions at each spaxel.  We thereby derive the RRL line width, velocity centroid, and intensity across the GCL. This method is discussed in more detail by \citet{2021ApJS..254...28A}. 

Using GDIGS RRL Gaussian fit parameters and the radio continuum map of \citet{law08}, we can derive a map of $T_{\rm e}$ toward the GCL following the method of \citet{quireza06b}. The RRL and continuum data products were both created using the GBT at similar center frequencies ($\sim 5.71$\,GHz for the RRL data and $\sim 4.85$\,GHz for the continuum data). Assuming that the gas is in local thermodynamic equilibrium (LTE),
\begin{multline} \left( \frac{T_{\rm e}^*}{\textnormal{K}} \right) = \left \{ 7103.3 \left( \frac{\nu _{\rm L}}{\textnormal{GHz}}\right)^{1.1} \left( \frac{T_{\rm C}}{T_{\rm L}(\textnormal{H}^+)} \right) \right. \\
\left. \left( \frac{\Delta V(\textnormal{H}^+)}{\textnormal{km\,s}^{-1}} \right)^{-1} \left( 1+y^+ \right)^{-1} \right \}^{0.87},\label{eq:etemp}
\end{multline}
where $\nu_{\rm L} = 5.7578$\,GHz is the average,
$T_{\rm sys}$-weighted, frequency of our Hn$\alpha$ recombination lines, 
$T_{\rm C}$ is the continuum antenna temperature, $T_{\rm L}$ is the H line antenna temperature, $\Delta V
(\textnormal{H}^+)$ is the hydrogen full width at half maximum (FWHM) line width, and $y^+ = 0.03$ is the average helium-to-hydrogen ionic abundance ratio of our field of view,
\begin{equation} y^+ = \frac{T_{\rm L}(^4\textnormal{He}^+)\Delta V (^4\textnormal{He}^+)}{T_{\rm L}(\textnormal{H}^+)\Delta V (\textnormal{H}^+)},
\label{eq:yplus}
\end{equation}
where $T_{\rm L}(^4\textnormal{He}^+)$ is the line temperature of helium and $\Delta V (^4\textnormal{He}^+)$ is the corresponding FWHM line width \citep{Peimbert1992}.  

The strong background continuum emission near the Galactic center makes it challenging to estimate the thermal continuum associated with the GCL directly. 
We estimate its intensity by fitting an exponential model of the form $T_{\rm C,\,bg} = a_1 e^{-a_2 b} + a_0$ to the portion of the radio continuum map outside the GCL, at $\ell \approx -0.9\degree$ and $|b|>0\degree$. In our model, $T_{\rm C,\,bg}$ is the background continuum temperature in K and $b$ is the Galactic latitude in degrees.  We find best fit values of $a_0 = -0.059 \pm 0.012$\,K, $a_1 = 2.145 \pm 0.026$\,K, and $a_2 = 4.730 \pm 0.127$\,deg$^{-1}$. 
We subtract the modeled background emission from the continuum data.

\begin{figure*}
    \centering
    \includegraphics[width=\textwidth]{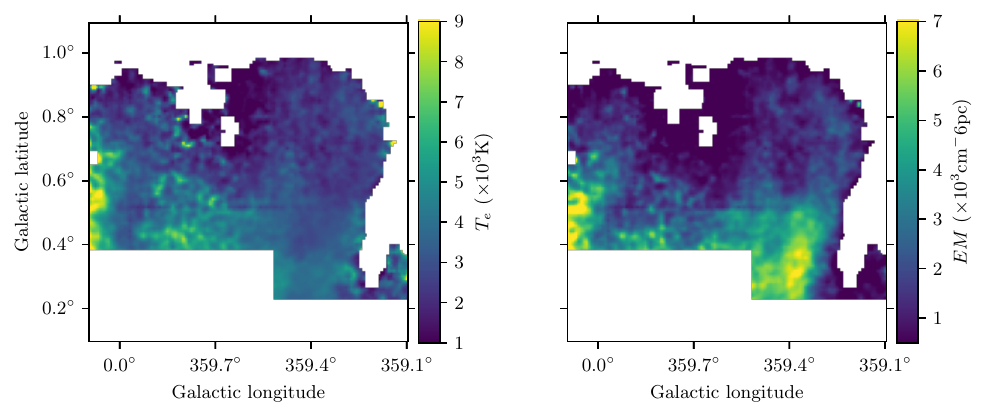}
\caption{Electron temperature $T_e$ (left panel) and emission measure $EM$ (right panel) derived using GDIGS RRL data and radio continuum data from \citet{law10}. Areas not used in the analysis (see Section~\ref{sec:gleam_abs}) are blanked. In areas of high signal-to-noise, $T_e\sim3{,}500\,\K$ and is fairly uniform, while $EM$ shows more variation, particularly with latitude.
    \label{fig:te}}
\end{figure*}

We calculate $T_{\rm e}$ at each spatial location using Equation~\ref{eq:etemp}, with the background-subtracted continuum, the RRL intensity, and the RRL line width maps as inputs. The resulting $T_e$ map is shown in the left panel of Figure~\ref{fig:te}.
The GCL electron temperatures range from 2{,}000 to 10{,}000\,\K, with larger $T_{\rm e}$ values found closer to the Galactic center and values of $T_e \simeq 3{,}500$\,K common for the locations of highest RRL intensity. 

For optically-thin Hn$\alpha$ GDIGS RRL emission,
\begin{equation}
\left( \frac{\rm EM}{\rm cm^{-6}\,pc} \right)
\simeq 
0.00981 \left(\frac{T_{\rm L}}{\K}\right) 
\left( \frac{\Delta V}{\kms} \right)
\left( \frac{T_{\rm e}}{\rm K} \right)^{3/2}\,,
\label{eq:em}
\end{equation}
where $EM$ is the emission measure, $T_L$ is the RRL amplitude, $\Delta V$ is the RRL FWHM, and $T_e$ is the electron temperature \citep[see Appendix of][]{2021ApJS..254...28A}. We use the intensity, line width, and $T_e$ maps to thereby determine an $EM$ map for the region, shown in the right panel of Figure~\ref{fig:te}.

\subsubsection{\textsc{Hii} Region G\,000.668+00.641\label{sec:G06}}
Using additional GBT RRL data, we derive the electron temperature $T_e$ and emission measure $EM$ of the \hii\ region G\,000.668+00.641.  This derivation is important for later analyses of the low-frequency radio emissivity (Section~\ref{sec:gleam_abs}).  There is H$\alpha$ emission from G\,000.668+00.641 (Figure~\ref{fig:gc}) and therefore it is likely to be within $\sim\!2$\,\kpc of the Sun.  Because of its location off the plane, G\,000.668+00.641 is relatively unconfused with other emission in the field.  

We perform position-switched RRL and continuum observations of G\,000.668+00.641 using the GBT at X-band (8-10\,\ghz), employing the same configuration and strategy as in \citet{anderson11, balser11}.  We observe RRLs on-source for a total of 9\,minutes at position $(\alpha, \delta$ [J2000]) = (17:44:42.58, $-$28:04:24.52) and off-source for the same amount of time, following the same track in the sky.  The on-source position is the location of peak intensity in the NVSS 1.4\,\ghz\ data \citep{condon98}.  We slew across G\,000.668+00.641 in continuum mode at a frequency of 8.665\,\ghz\ in scans of constant RA and Dec.  In total we make 10 scans across the source.  The system temperature during observations was $\sim\!45\,\K$.  Calibration via noise diodes fired during observations should be good to within 10\% \citep{balser11}, although a calibration uncertainty will affect the line and continuum data similarly.

We average all seven RRLs in the bandpass (H87$\alpha$ to H93$\alpha$), at two polarizations, smooth to 1.9\,\kms\ velocity resolution, remove a polynomial baseline, and fit Gaussian components. The hydrogen RRL data are best described by a single Gaussian of height $T_L({\rm H^+}) = 58.5\pm0.2\,\mK$ and FWHM $\Delta V({\rm H^+}) = 20.1\pm0.1\,\kms$.  The corresponding helium RRL fit is $T_L(^4{\rm He^+}) = 6.0\pm0.4\,\mK$ and FWHM $\Delta V(^4{\rm He^+}) = 13.0\pm0.9\,\kms$.  The rms noise in the data is 2.816\,mK, making these detections $47\sigma$ and $4.8\sigma$, respectively \citep[cf.][]{lenz92}.  The peak of hydrogen RRL emission is at $2.8\pm0.1\,\kms$ \citep[cf.][3.7\,\kms]{lockman96}.

We average all continuum scans from a given direction and remove a polynomial baseline.  The RA scans have a simple profile that is easily reproduced by a single Gaussian of height $352\pm5\,\mK$ and FWHM $2.51\pm0.05\arcmin$.  The Dec scans are more complicated, and a Gaussian fit to the peak of emission averages $372\pm20\,\mK$ and FWHM $2.06\pm0.07\arcmin$.  Scans from both directions peak within $0.2\arcmin$ of the targeted source position.  The baseline is the largest source of uncertainty in the continuum data.  Because the RA scan profile is much simpler than that of the Dec scan, we believe that the derived RA-scan parameters are closer to the true intensity, but we conservatively take the Dec-scan uncertainty.  Our final continuum intensity is therefore $T_C =  350\pm20\,\mK$.

Using Equation~\ref{eq:etemp}, and scaling our value down by 5.7\% to account for the frequency difference between line and continuum observations \citep{wenger19}, we find an electron temperature for G\,000.668+00.641 of $T_e = 5700\pm 570\,\K$.  The emission measure is therefore $EM=8\times10^3$\,cm$^{-6}$\,pc.
From Equation~\ref{eq:yplus}, we find a value of $y^+ = 0.066\pm0.013$, where the 20\% uncertainty is estimated from different baseline fits.  

We estimate a 10\% uncertainty on $T_e$.  This estimate does not include calibration uncertainties since they will not strongly affect the ratio $T_C/T_L$.  The continuum intensity is conservatively uncertain by about 10\% due to uncertainty in determining the zero-point offset, but the RRL intensity and line widths have low uncertainties.  The uncertainty of $y^+$ we take as 20\%, but this results in only a few percent uncertainty in $T_e$.

\section{Nature of the GCL Radio Emission\label{sec:gleam_abs}}

Only synchrotron self-absorption (SSA) and free-free absorption (FFA) can explain the low-frequency absorption of the GCL seen by GLEAM (Figure~\ref{fig:gc}); thus, we address each in turn.

\subsection{Synchrotron self-absorption}

SSA occurs when the brightness temperature of a radio source approaches the electron temperature, causing the source to become opaque. A self-absorbed synchrotron source produces brightness (and electron) temperatures of $\sqrt{1.4 \times 10^{12} (\frac{\nu}{\mathrm{Hz}}) B^{-1}}$, where $\nu$ is the turnover frequency.

By extrapolating from measurements of nearby regions, 
\cite{heywood19} found that the MeerKAT 1.2\,\ghz\ radio emission of the GCL is produced by synchrotron, but they were unable to determine the spectral index from their data due to the confusing nature of the region. \cite{law09} analysed GBT data at 0.33, 1.5, 5, and 8.5\,\ghz\ and also concluded that the radio continuum emission associated with the GCL is dominated by synchrotron, and that its magnetic field strength is of order 50\,$\mu$G.  These studies would predict that the radio continuum absorption is due to SSA.

The GLEAM data show clear absorption below $\nu=100$\,\mhz, so using this frequency value and $B=50$\,$\mu$G, we obtain a brightness (and electron) temperature of $1.7\times10^{12}$\,K.  This value is clearly not observed; we therefore conclude that the absorption must be FFA.

\subsection{Free-free absorption}

The observed brightness temperature, $T_B$, towards a thermally-emitting source is:
\begin{equation}
  T_B =T_e(1-e^{-\tau(\nu)})+T_{gb}e^{-{\tau(\nu)}} +T_{gf},
\label{eq:temperature}\end{equation}
where $T_e$ is the electron temperature, $T_{gb}$ is the temperature of the non-thermal Galactic background, $T_{gf}$ is the temperature of the non-thermal Galactic foreground, and $\tau(\nu)$ is the optical depth.  Structures dominated by the emission from thermal electrons become opaque ($\tau \geq 1$) at frequencies below $\sim\!100\,\mhz$, with the optical depth given by:
\begin{equation}
\tau(\nu) \simeq 8.2\times 10^{-2} \, \left(\frac{T_e}{\K}\right)^{-1.35}\, \left(\frac{\phantom{T}\nu\phantom{T}}{\ghz}\right)^{-2.1} \left(\frac{{ EM}}{\cm^{-6}\,\pc}\right)\,,
\label{eq:tau}\end{equation}

\begin{figure}
    \centering
    \includegraphics[width=\linewidth]{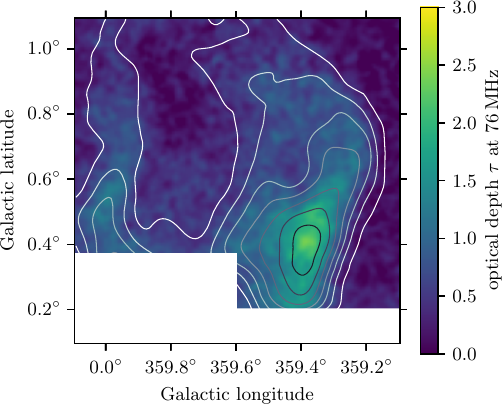}
    \caption{The optical depth at 76\,MHz, calculated via Equation~\ref{eq:tau}, using the EM and $T_e$ shown in \Fig~\ref{fig:gcl_emi}. The 76-MHz GLEAM brightness temperature decrement data is overlaid using 8~contours of levels $-30$ to $-2\times10^3$\,K , showing very similar morphology. Areas where the background subtraction of the GLEAM data was of poor quality are blanked.}
    \label{fig:tau}
\end{figure}

The calculated optical depth at 76\,MHz is shown in \Fig~\ref{fig:tau}. The brightness temperature decrement at 76-MHz, derived independently from the GLEAM data, is overlaid. The two maps show strong morphological similarity indicating a common origin.

From Equation~\ref{eq:tau} we expect thermal absorption to produce an exponential drop-off in radio brightness as the optical depth becomes large (at lower frequencies). We can compare the brightness temperature of the GCL as a function of frequency to that of nearby thermal and synchrotron-emitting objects.
We select two regions of the GCL that have high radio continuum signal-to-noise (yellow boxes in bottom right panel of Figure~\ref{fig:sobel}) and measure the average temperature as a function of frequency across the 20 background-subtracted GLEAM sub-band images.  We compute the uncertainties at each sub-band from the standard deviation in the pixel values of a nearby region free of discrete sources of emission, (dashed yellow box in top left panel of Figure~\ref{fig:sobel}), added in quadrature with a 2\% calibration uncertainty
\citep{2017MNRAS.464.1146H} and a 5\% uncertainty estimated from the background subtraction process.

Using Equations~\ref{eq:temperature} and \ref{eq:tau}, we derive the predicted brightness temperature as a function of frequency using the measured temperature decrement in one sub-band, a measurement of the emission measure, and an electron temperature.  We derive the latter two quantities in Section~\ref{sec:G06} for the \hii\ region G\,000.668+00.641. We use the 76\,\mhz\ GLEAM sub-band data to derive the reference temperature decrement. This sub-band has the largest optical depth, so the effect of measurement errors on the derivation of the temperature profile will be minimized (since $\Delta T \propto e^{-\tau(\nu)}$).

\begin{figure*}
    \centering
    \includegraphics[width=2.3in]{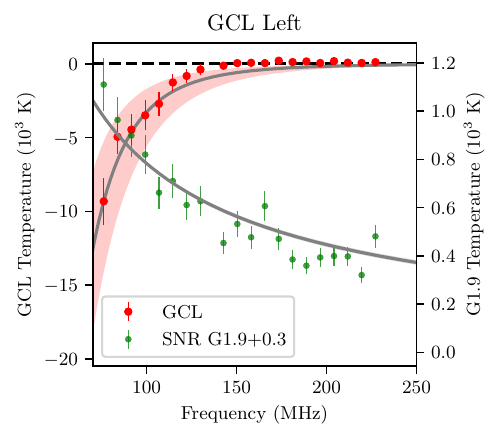}
    \includegraphics[width=2.3in]{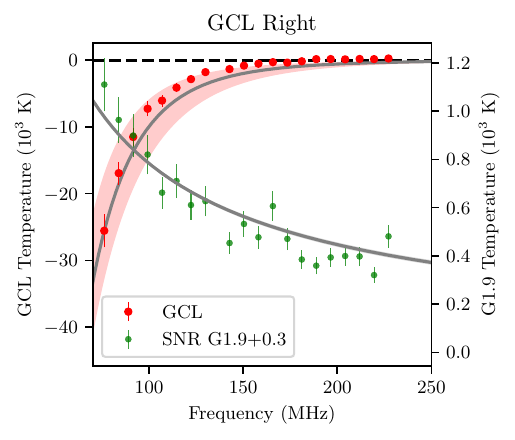}
    \includegraphics[width=2.3in]{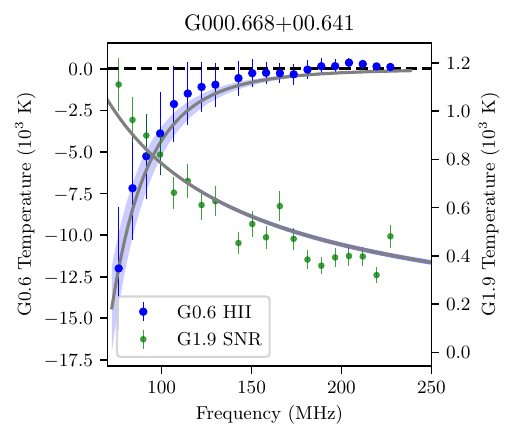}
    \caption{
    Temperature profiles as a function of frequency.  The left panel shows the profile for the GCL region identified in the left yellow box in the bottom right panel of Figure~\ref{fig:sobel}; the middle panel shows the profile for the right yellow box in the same figure; the right panel shows the profile for the \hii\ region G\,000.668+00.641. The red/blue points show measured temperatures from the 20 background-subtracted GLEAM sub-bands. The grey curves show predicted temperature profiles using the 76\,\mhz data and the $EM$ and $T_e$ values from the \hii region G\,000.668+00.641 (it is not a fit to the points). The shaded red/blue regions show the 1-$\sigma$ range of profile predictions given the uncertainties on the input quantities.  The green points are of the SNR G\,1.9+0.3 \citep{2020MNRAS.492.2606L}, with their derived fit as the gray curve. The SNR shows a spectrum typical of non-thermal synchrotron emission, while both lobes of the GCL and the comparison \hii\ region show flat spectra with exponential roll-offs typical of thermal emission and free-free absorption.
    \label{fig:SEDs}}
\end{figure*}

In Figure~\ref{fig:SEDs} we plot the temperature (intensity) as a function of frequency for the two selected regions in the GCL (see Figure~\ref{fig:sobel}), with the predicted temperature curve overlaid (not fit). 
This shows that the low-frequency emission of the GCL is characteristic of a thermally-emitting source, and not consistent with that of a non-thermal source.
There is good agreement between the predicted profile and the temperature measurements. We also show the derived GLEAM brightness temperatures and the power-law fit of the nearby SNR G\,1.9+0.3 from \citet{2020MNRAS.492.2606L}. The power-law brightness distribution is
typical of the synchrotron emission expected from a supernova remnant. 

We perform a similar GLEAM background-subtraction to that shown in Figure~\ref{fig:sobel}, and derive the temperature profile of the \hii\ region, G\,000.668+00.641, shown in the third panel of Figure~\ref{fig:SEDs}.  These data are very similar to those of the GCL. \cite{2016PASA...33...20H} analyzed GLEAM data to show the temperature profiles of a 61~\hii\ regions, and their low-frequency radio emission profiles are similar to those of G\,000.668+00.641 and the GCL. These comparisons show that the low-frequency emission of the GCL is consistent with that of an \hii\ region and not consistent with that of a synchrotron emitting object.

Based on the similarity of the GCL temperature profiles to those of G\,000.668+00.641 and the \hii\ regions in \citet{2016PASA...33...20H}, and the dissimilarity with the temperature profile of SNR G\,1.9+0.3, we conclude that the low-frequency emission and absorption features of the GCL are produced by thermal electrons. 

\section{Distance to and emissivity toward the GCL}

Thermal absorption of low-frequency radio continuum emission can be used to determine distances using the measured emissivity of the source (in \K\,pc$^{-1}$) and the characteristic emissivity along the line of sight. Following Equation~\ref{eq:temperature}, if $T_{gf}$ remains relatively constant over the field-of-view, one can derive the brightness temperature of the Galactic background by measuring $\Delta T$, the difference between the temperature on-source ($T_B$) and the temperature of the background ($T_{gb}$):
\begin{equation}\label{eq:deltaT}
\Delta T = \left(T_e - T_{gb}\right) \left(1 - e^{-\tau (\nu)}\right)
\end{equation}
One can then derive the emissivity $\epsilon$ behind the absorbing source:
\begin{equation}\label{eq:emi}
\frac{\epsilon}{\K\pc^{-1}} = \frac{1}{1000}\left(\frac{T_{gb}}{\K}\right) \left(\frac{d_\mathrm{edge}}{\kpc}\right)^{-1}\,,
\end{equation}
where $d_\mathrm{edge}$ is the distance between the \hii\ region and the ``edge'' of the Galaxy.
This technique was employed by \cite{2006AJ....132..242N} using 74\,\mhz\ VLA observations of the inner Galaxy, and more recently by \cite{2017MNRAS.465.3163S, 2018MNRAS.479.4041S}
using MWA observations of a large population of Galactic \hii\ regions.  The method assumes that $T_e$ and $EM$ do not vary over the extent of the source, and that $\epsilon$ does not drastically change on angular scales smaller than the source.

\begin{figure}
    \centering
    \includegraphics[width=\linewidth]{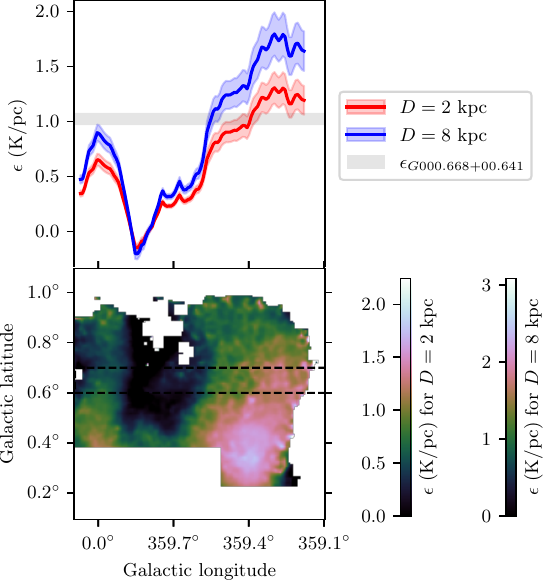}
    \caption{The derived emissivity $\epsilon$ over the GCL, using our measured values of $T_e$, $EM$, and $\Delta T$. The bottom-left panel shows $\epsilon$ over the region where our measurements are reliable. The associated colorbars show the values of $\epsilon$ for two different distance assumptions: local (2\,kpc) and at the Galactic Centre (8\,kpc). The top panel shows the average profile of the emissivity over the region $0.591^\circ < b < 0.691^\circ$, i.e. the same latitude as G\,000.668+00.641.
    \label{fig:gcl_emi}}
\end{figure}

We calculate the emissivity $\epsilon$ over the extent of the GCL for heliocentric distances of 2 and 8\,\kpc\ by substituting $T_{gb}$ from Equation~\ref{eq:deltaT} into Equation~\ref{eq:emi}, and using our measured values of $T_e$, $EM$, and $\Delta T$ (see Figure~\ref{fig:gcl_emi}).  We compare our GCL emissivity measurements to those of the nearby \hii\ region G\,000.668+00.641, which might be expected to have a similar background emissivity. For G\,000.668+00.641, $T_e=5700\pm570\,\K$ and $EM=8\times10^3$\,cm$^{-6}$\pc (Section~\ref{sec:G06}), and therefore $\tau(\mathrm{76\,\mhz}) = 1.25$. Based on the presence of H$\alpha$ emission (Figure~\ref{fig:gc}), we assume a distance to G\,000.668+00.641 of $d=2\pm1$\,kpc (i.e. $d_\mathrm{edge}=22\pm1$\,\kpc\ for a Galactic radius of 12\,\kpc) and measure the 76\,\mhz\ temperature difference to be $\Delta T=-12000\pm960$\,\K
These values yield a background 76\,\mhz\ emissivity of $\epsilon=1.1\pm0.1$\,K\,pc$^{-1}$, which we show as the shaded grey horizontal bar in the top panel of Figure~\ref{fig:gcl_emi}.

The emissivity varies strongly over the GCL, with a general increase in $\epsilon$ with decreasing Galactic latitude. There is also a difference in the typical $\epsilon$ between the left and right ``lobes'' of the GCL (i.e $l\approx0\degree$ and $l\approx359.3\degree$). There is no clear agreement with the background emissivity calculated from G\,000.668+00.641, even for the same Galactic latitude range (top panel of Figure~\ref{fig:gcl_emi}). 

We therefore conclude that the background emissivity varies too much for this technique to place a strong constraint on the distance to the GCL. We note, however, that the values of $\epsilon$ derived for the lower distance ($d=2$\,kpc) are more consistent with the range $0.3<\epsilon<1.0\,\K\,{\rm pc}^{-1}$ derived using LOFAR \citep{2019A&A...621A.127P} and the VLA \citep{2006AJ....132..242N}.

\section{Summary}

This is one of two papers which presents new evidence on the nature and distance of the GCL. Here we have shown that:

\begin{itemize}
    \item The GCL has a strong turnover in the $\sim\!70-200\,\mhz$ GLEAM data, and this cannot be explained by synchrotron self-absorption, only free-free absorption;
    \item Using RRL and radio continuum data, we measure a low and relatively consistent electron temperature across the GCL of $\lsim 3500\,\K$;
    \item Significant variations in emissivity are measured over the GCL on scales of $\sim$0.1$^\circ$;
    \item Either the background emissivity is enhanced by $\sim$75\,\% toward the GCL, or it has a distance smaller than 8\,kpc, with a distance of 2\,kpc being most consistent with the measurements, and other previous work.
\end{itemize}

In the companion paper we show that the GCL has infrared and radio properties similar to those of Galactic \hii regions.

\begin{acknowledgments}
NHW is supported by an Australian Research Council Future Fellowship (project number FT190100231) funded by the Australian Government. 
\nraoblurb\   We thank West Virginia University for its financial support of GBT operations, which enabled some of the observations for
this project.  This work is supported by NSF grant AST1516021 to LDA,  NASA grant NNX17AJ27GRAB to RAB, and Australian Research Council (ARC)  grant DP160100723 to NM-G.  Some of this work took part under the program SoStar of the PSI2 project funded by the IDEX Paris-Saclay, ANR-11-IDEX-0003-02. This research has made use of the SIMBAD database, operated at CDS, Strasbourg, France.  RAB would also like to acknowledge useful conversations on extinction distances with Gregory Green and Doug Finkbeiner.  
\end{acknowledgments}

\bibliographystyle{apj}
\bibliography{ref.bib}

\end{document}

%% file: preamble.tex
\usepackage{xspace}
\usepackage{graphicx}
\usepackage{natbib}
\usepackage{amsmath}
\usepackage{booktabs}
\usepackage{multirow}
\usepackage{afterpage}
\usepackage{rotating}

\newcommand{\nraoblurb}{The National Radio Astronomy Observatory is
a facility of the National Science Foundation operated under cooperative
agreement by Associated Universities, Inc.}
\newcommand{\hide}[1]{}

\newcommand{\lsim}{\ensuremath{\,\lesssim\,}\xspace}
\newcommand{\kms}{\ensuremath{\,{\rm km\,s^{-1}}}\xspace}

\newcommand{\cm}{\ensuremath{\,{\rm cm}}\xspace}
\newcommand{\pc}{\ensuremath{\,{\rm pc}}\xspace}
\newcommand{\kpc}{\ensuremath{\,{\rm kpc}}\xspace}
\newcommand{\K}{\ensuremath{\,{\rm K}}\xspace}
\newcommand{\mK}{\ensuremath{\,{\rm mK}}\xspace}

\newcommand{\mhz}{\ensuremath{\,{\rm MHz}}\xspace}
\newcommand{\ghz}{\ensuremath{\,{\rm GHz}}\xspace}

\newcommand{\degree}{\ensuremath{^\circ}\xspace}

\newcommand{\jyb}{\ensuremath{{\rm \,Jy\,beam^{-1}}}\xspace}

\newcommand{\msun}{\ensuremath{\,M_\odot}\xspace}     
 
\newcommand{\hii}{{\rm H\,{\footnotesize II}}\xspace}

\newcommand{\hna}{\ensuremath{{\rm H}\,{\rm n}\,\alpha\xspace}}

\newcommand{\Fig}{Fig.}